# The Age-related Differences in Web Information Search Process

Zhaopeng Xing [1][0000-0002-3982-4299], Xiaojun (Jenny) Yuan[2], Lisa Vizer[1]

[1] University of North Carolina at Chapel Hill, Chapel Hill, USA

[2] University at Albany - State University of New York, Albany, USA

**Abstract.** Older adults' need for quality health information has never been more critical as during the COVID-19 pandemic. Yet, they are susceptible to the widespread misinformation disseminated through search engines and social media. To build a search-related behavioral profile of older adults, this article surveys the empirical research on age-related differences in query formulation, search strategies, information evaluation, and susceptibility to misinformation effects. It also decomposes the mechanisms (i.e., cognitive changes, development goal shift) and moderators (i.e., search task and interface design) of such differences. To inform the design of information systems to improve older adults' information search experience, we discuss opportunities for future research.

## 1 Introduction

Older adults as a cohort are at higher risk of infection and death due to COVID-19 [1]. and are at a disadvantage in accessing quality health information. Although the volume of information regarding COVID-19 is exploding, it is rarely found on trusted websites [2] while misinformation—false statements and claims created and distributed unintentionally [3]—is widely disseminated through web searches and social networks [2, 4, 5]. Since older adults are more susceptible to this misinformation [6], this can lead to false beliefs and, in turn, poor decision-making regarding health care [7, 8]. With the impending second wave of COVID-19 [9], researchers in information science and human-computer interaction have an opportunity to facilitate older adults' information access through adaptive search interfaces that are sensitive to age-related differences in search behavior and minimize the discrepancy in misinformation effects.

Age-related differences in search behavior are well-characterized from information search studies. For example, older adults generally have lower global search performance than their younger counterparts. They spend a longer time on search tasks, exhibit lower completion speed, and retrieve fewer correct or accurate search results [10–15]. Older adults also show differences with respect to query formation, search strategy, and information evaluation (e.g., [11, 16–19]). Despite the evidence, there is still a lack of an integrated understanding of these different search patterns and the mechanisms behind them. [20] proposed a conceptual framework that integrates the research regarding the age-related differences in the decision-making process. It further specifies the influences of aging mechanisms and environmental factors related to experimental contexts and task characteristics. However, it interprets information search process simply as an actionable stage and thus lacks an "information interaction" lens into age-related difference. That the information search in the digital information environment is an iterative, multi-step and interactive process has long been accepted [21–23]. In this process, a searcher's information needs are fulfilled in a natural and efficient interaction with an information retrieval (IR) interface which considers the searcher's requests and context in order to present appropriate answer [24]. This literature survey uses this information interaction lens as a means to better understand age-related differences in the search process and make recommendations to facilitate the inclusive design of IR systems for older adults. In addition, the impact of misinformation effects is addressed in this survey, shedding light on the role of misinformation in the context of age-related differences in search in order to identify opportunities for future research. This article synthesizes prior empirical evidence that characterizes age-related differences in information search interaction in terms of query formulation, search strategies, information evaluation, and susceptibility to misinformation effects. We also discuss the mechanisms underlying age-related differences, including cognitive changes, development goal shifts, search tasks, and interface design.

## 2 Age-related difference in search behavior

### 2.1 Query formulation

Older adults demonstrate different query formulation patterns than younger people. They are less likely than their younger counterparts to issue successful and topically relevant queries or employ appropriate query strategies. Specifically, during a search they issue fewer queries and use fewer keywords in those queries [11, 12, 18, 25], often resulting in queries that fail to retrieve the relevant information [10, 11, 26, 27]. As a search proceeds, older adults are less likely to use common query reformulation tactics, such as adding or removing words, and the semantic relevance of their queries is significantly reduced as well [12, 25]. When they encounter query formulation challenges, older adults tend to resort to using keywords from task descriptions for external support [11, 13].

### 2.2 Search strategy

When processing information, older adults use different strategies in allocating attentional resources than younger searchers. They generally employ a simpler and less cognitively demanding search strategies that minimize their interaction with information and systems [14, 16, 28]. In particular, older adults favor a knowledge-based search strategy where they spend a longer time examining and interpreting results based on prior knowledge rather than searching, planning, and regulating the search process [10, 11, 17, 26]. They also tend to utilize a search strategy that is more exploitative and less explorative. That is, they make few changes to the query, visit fewer results links, and exhaustively navigate through webpages linked to a single visited result [29, 30]. However, regardless of search strategy, older adults are comparable to younger searchers in their ability to adapt their strategy to the demands of a search task [14, 17, 29, 31, 32].

### 2.3 Information evaluation

Researchers have also noted age-related differences in result evaluation, especially regarding health care information (see a review [33]). Older adults have a lower level of trust in online information sources and prefer to obtain health information through interpersonal communication. They also show a lower level of skill in judging the quality and reliability of online information as well as lower sensitivity to the credibility cues in content (i.e., evidence, argument rigor, information quality and bias) and interface features (i.e., design look and source indicators) [34]. Older adults rely on a more limited set of indicators to assess website information quality and associate complicated information with credible information [35]. Furthermore, older adults' lower familiarity with the internet translates to unclear criteria for evaluating and classifying online information sources (public vs. commercial) [36]. This evaluation of online information is a primary challenge for older adults [27].

In addition to challenges with evaluating information quality, older adults also demonstrate a significant bias towards positive information, known as the positivity effect. This effect emerges in the information search stage of a decision-making process (see a meta-review [37]) when older adults exhibit greater attention to and memory for positive information when assessing options [38, 39]. Although the preference for positive information can lead to more positive post-decisional emotion and higher decision-making satisfaction, this effect can bias the judgment of results and, in turn, the quality of decisions [40].

### 2.4 Susceptibility to misinformation effects

Older adults' lower ability to evaluate online information and their preference for positive information may increase their susceptibility to misleading information [6]. Misinformation can influence search outcomes [8] exacerbate mental and emotional stress (see a review [3]), or distort memories of an event—all part of the misinformation effect [41] to which older adults are vulnerable [6]. Older adults experience difficulty in recognizing misinformation and recalling original information [42, 43]. They are also more confident in false memories and less confident in accurate memories [44]. However, when prompted with cues about the reliability of a source, the age-related differences in evaluating misinformation are reduced, indicating that older adults need more external support to recognize misinformation.

## 3  Age-related difference in search behavior

### 3.1  Cognitive changes

Age-related differences in search behavior are extensively attributed to normal age-induced changes in cognition. During the aging process, fluid and crystallized intelligence change [45], and those changes impact performance on search tasks. *Fluid intelligence (Gf)* is the ability to adapt to unusual or novel problems, and employs reasoning, categorizing, information processing, and working memory capabilities [46, 47]. When executing an information search, fluid intelligence supports decision making, the transition between search sessions, and adaptation of unsuccessful searches. This series of events relies on working memory to update the mental representation of information needs by replacing less relevant information, inhibition to suppress or inhibit irrelevant information, and cognitive flexibility to switch attentional resources between different searches [48]. These abilities decline with aging and the decline is associated with the reduced search performance, challenges with query formulation [11, 17, 49, 50], and lower ability to evaluate information quality [51]. *Crystallized intelligence (Gc)*, on the other hand, reflects the prior experience and acquired knowledge derived from education and acculturation [46, 47] and increases throughout adulthood [45]. Higher verbal and domain knowledge helps older adults perform better in web navigation and query reformulation [11, 50, 52] and expedites the search and decision-making process [53]. Although knowledge-based strategies also help integrate prior knowledge with information gained from a search [17], older adults' generally lower familiarity with computers and the internet can negatively impact their ability to evaluate information credibility. [36, 54] also found an association between lower knowledge of technology with the lower trust of website contents.

Older adults' selection of search strategy partially reflects an economic trade-off between fluid and crystallized intelligence. They often rely on prior knowledge in the search process and allocate more attention to a narrow but deep results space to avoid frequent switches between different strategies and between information. The purpose of this strategy is to mitigate declines in fluid intelligence, maximize search outcomes, and reduce cognitive cost [17, 29, 31, 32].

Regarding the impact of age-related cognitive changes on the effect of misinformation, there have been two neurocognitive explanations. [55] attribute increased susceptibility to the decline of the source-monitoring ability. This ability is associated with the function of the brain's frontal lobe, which primarily supports decision making and controls the selection and coordination of goal-directed behavior. It retrieves contextual information to assist with judging the source of remembered information [42, 56, 57]. Another explanation lies in age-related differences in the mental presentation of information. Global presence preserves the general meaning of information while a specific presence preserves precise features of information [58, 59]. Older adults tend to rely more on global presentation and thus leave out details about original information [60].

### 3.2 Development goal shift

*Motivation.* During the aging process, individual development goals get shifted. A shift takes place from a "striving" pattern, focused on influencing the external environment, to a "preservative" pattern, focused on avoiding losses and failures by selectively engaging in goals. This shift, known as the lifespan theory of control [61, 62], is associated with cognitive deficits and reflected in older adults' choice of adaptive strategies for allocating motivational resources in response to challenging tasks. Therefore, it may partially explain older adults' inclination to use knowledge-based and exploitative search strategies and their reliance on external assistance to formulate queries.

*Emotion.* Older adults also adjust emotion regulation strategies to prioritize goals that benefit affect wellness and close relationships [63, 64]. This results in a tendency to attend to positive stimuli over negative stimuli, promoting positive emotional experiences and avoiding negative ones [37, 65]. This tendency shows up in older adults' preference towards positive search results and can predispose older adults to believe positive information and avoid important negative information during a pandemic. A preference for positive emotion may also strengthen the age-related discrepancy in misinformation effects. Since mood can influence information processing style [66], and positive moods are associated with general and relational information processing approaches, they can increase failures to retrieve contextual information and false memories [43].

### 3.3 Search tasks

Older adults' reactions to environmental factors are also different from younger people. Task complexity and older adults' perception of task relevance moderate age-related search differences. Older adults are able to adapt to task factors. [17] investigated task complexity using well-defined versus ill-defined tasks where participants needed to view either single or multiple webpages to obtain answers. Older adults performed better than younger people on ill-defined tasks because of the advantages in knowledge-based search strategies. Further study also found their adaptions to low accessible tasks (i.e., the tasks require more search keywords) by reducing their website visits [29]. However, they still encounter more challenges in complex tasks. In the search tasks that are hard or impossible to find results, older adults have more difficulties in getting out of the impasses by reformulating the queries [10, 11, 26]. Their search accuracy and semantic relevance of queries significantly drops when the overlaps between the task description and targeted webpage content decreases [25] The task with ambiguous search goals, which increasingly requires one's ability to infer queries, are also more challenging and cognitive taxing for older adults than the young [13, 52].

Early evidence has also shown that older adults are more sensitive to the perceived relevance of a task, which increases the salience of the self-assessed importance of a task and maximizes the cost-efficiency of cognitive engagement [28]. For tasks they perceive to be highly relevant, older adults utilize a more systematic and complex search strategy, whereas, for tasks they deem to be of lower relevance, they use a "good enough" strategy [67]. This adaptation is associated with the use of a non-compensatory strategy, where a searcher evaluates a subset, instead of the full set, of the targeted information attributes when assessing results [68]. This results in a locally optimal rather than a globally optimal outcome. The adaptive process diminishes age-related discrepancies in search accuracy with little impact on decision quality [16, 31]. Also, this sensitivity to highly relevant information reduces the age-related differences resulting from the positivity effect. For example, older adults in poorer health, who tend to be more sensitive to have relevance health-related information, might show less of a positivity effect when seeking health-related information [40].

### 3.4 Interface design

Interface design determines how information is presented and effective design can moderate age-related differences. Tag-based IR interfaces that organize information according to keywords in a flat structure can help older adults take advantage of crystallized intelligence and reduce the cognitive demands associated with perception and working memory [50]. The web interfaces with poor ergonomics exacerbate age-related declines and lead to longer search times and a more complex search process [15]. Older adults are also less adaptable to the differences in interface design. They are somewhat resistant to adjusting their search strategies to align with a website's information structure and are less impacted by differences in interface layouts [32]. However, certain functionality in search interfaces is still useful. It can remind older adults of their search goals during the search process, which has been proved to improve older adults' query attention allocation, formulation efficiency, and search strategy adaption [69]. Despite evidence that older adults are less sensitive to some interface features than the younger [34], they do associate elements of good interface design, such as clear layout and interactive features, with higher website credibility [54].

## 4 Discussion

Age-related declines notwithstanding, older adults can employ effective adaptive strategies to mitigate those declines and maximize informational and emotional gain. They can also execute complex search tasks by taking advantage of vocabulary or domain knowledge. However, they do experience challenges and the design of IR systems should take into consideration older adults' unique search patterns to support search performance and user experience. For example, spoken conversational interfaces (SCIs) may assist older adults in generating efficient queries and help uncover information needs. Although its potential in engendering a more natural and interactive search experience, few studies have examined the influences of SCIs on older adults' search behavior and the moderate effects of those age-induced changes in the conversational search interaction (see a review [70]). By enabling a conversational search paradigm, SCIs may introduce design changes and functionality to improve cognitive engagement, support adaptive search strategy, reduce positivity effects, and highlight topics a user perceives as highly relevant. In addition, further support is also needed to help older adults identify credible information and diminish misinformation effects.

Strategies might include assistance with developing information evaluation skills [71] and incorporating input from trusted third parties such as family members, close friends, and healthcare professionals [72]. To further educate searchers on how to evaluate the credibility of search results, search interfaces should clearly label each site's credentials and content sources.

The current approaches to facilitating information search for older adults usually attempt to activate aspects of cognitive intelligence, such as working memory [69], or leverage prior knowledge, as when introducing external assistance [50]. However, few systems seek to help improve older adults' capabilities. Research might investigate how to enhance older adults' capabilities through innovations such as a symbiotic IR system [73]. [74] proposes a Human Engaged Computing theoretical framework where a system supports self-efficacy by helping develop the users' skills and abilities rather than performing tasks for them. Further study is needed to understand how best to apply such frameworks to the design of IR systems.

Misinformation research has commonly fallen under the purview of neuroscience, law, political science, and social journalism. However, the COVID-19 pandemic has made clear that although access to trustworthy information is important for everyone, it is particularly crucial for older adults and should be better supported by IR systems. Future research should examine how older adults acquire, recognize, and eliminate health-related misinformation in the context of interactive IR and how that process might be better supported by IR systems, especially in the context of the COVID-19 pandemic. It is also important to examine the potential health impacts of misinformation and how to remediate those impacts with advanced IR systems.

## 5        Conclusion

This survey synthesizes the evidence on age-related differences in information search, the mechanisms behind those differences, and the contextual factors that moderate differences. Older adults use adaptive search strategies to mitigate age-related declines, but challenges are still apparent. Future research directions should investigate age-related differences in conversational search, effective interface design, improved access to credible information, symbiotic IR systems, and age-related differences in the impact of misinformation. The COVID-19 pandemic lends increased urgency to these investigations as misinformation threatens older adults' ability to easily access trusted health information in the course of a public health emergency.